\documentclass{llncs}
\usepackage{url}
\usepackage{color}
\usepackage{paralist}
\usepackage{soul}
\usepackage{mathtools}
\DeclarePairedDelimiter\parens{\lparen}{\rparen}
\usepackage{array}
\usepackage{booktabs}
\usepackage{graphicx}
\usepackage{multirow}
\newcolumntype{L}[1]{>{\raggedright\let\newline\\\arraybackslash\hspace{0pt}}m{#1}}
\newcolumntype{C}[1]{>{\centering\let\newline\\\arraybackslash\hspace{0pt}}m{#1}}
\newcolumntype{R}[1]{>{\raggedleft\let\newline\\\arraybackslash\hspace{0pt}}m{#1}}
\usepackage{tabu}
\usepackage{subfigure}

\newcommand{\rlzap}
	{\textsf{RLZAP}}
\newcommand{\rlz}
	{\textsf{RLZ}}
\newcommand{\rank}
	{\ensuremath{\mathsf{rank}}}
\newcommand{\select}
	{\ensuremath{\mathsf{select}}}
\newcommand{\BWT}
	{\ensuremath{\mathsf{BWT}}}
\newcommand{\LCP}
	{\ensuremath{\mathsf{LCP}}}

\newcommand{\MinLength}
	{\ensuremath{\mathsf{MinLen}}}

\newcommand{\LookAhead}
	{\ensuremath{\mathsf{LookAhead}}}

\newcommand{\DeltaBits}
	{\ensuremath{\mathsf{DeltaBits}}}
\newcommand{\MinExplicitLength}
	{\ensuremath{\mathsf{Explicit}_{\mathsf{Len}}}}
\newcommand{\MaxLit}
	{\ensuremath{\mathsf{MaxLit}}}
\newcommand{\SampleInterval}
	{\ensuremath{\mathsf{SampleInt}}}
\newcommand{\MatchRelPtr}[1]
	{\ensuremath{\mathsf{MatchPtr\parens{#1}}}}
\newcommand{\MatchLen}[1]
	{\ensuremath{\mathsf{MatchLen\parens{#1}}}}
\newcommand{\NExc}
	{\ensuremath{\mathsf{Exc}}}
\newcommand{\PhraseBV}
	{\ensuremath{\mathsf{P}}}
\newcommand{\ExplicitBV}
	{\ensuremath{\mathsf{E}}}

\newcommand{\Rel}[1]
	{\ensuremath{\mathsf{Rel\parens{#1}}}}
\newcommand{\Start}[1]
	{\ensuremath{\mathsf{Start\parens{#1}}}}
\newcommand{\Length}[1]
	{\ensuremath{\mathsf{Len\parens{#1}}}}
\newcommand{\Literals}[1]
	{\ensuremath{\mathsf{Lits\parens{#1}}}}
\newcommand{\PrevPhr}[1]
	{\ensuremath{\mathsf{Prev\parens{#1}}}}
\newcommand{\AbsPhr}[1]
	{\ensuremath{\mathsf{Abs\parens{#1}}}}
\newcommand{\Prefix}
	{\ensuremath{\mathsf{Prefix}}}
\newcommand{\StaticTable}
	{\ensuremath{\mathsf{Byte}_\Sigma}}

\begin{document}

\title{RLZAP:\\Relative Lempel-Ziv with Adaptive Pointers\thanks{Supported by the Academy of Finland through grants 258308, 268324, 284598 and 285221.  Parts of this work were done during the second author's visit to the University of Helsinki and during the third author's visits to Illumina Cambridge Ltd.\ and the University of A Coru\~na, Spain.}}
\author{Anthony J.\ Cox\inst{1} \and
  Andrea Farruggia\inst{2} \and
  Travis Gagie\inst{3,4} \and\\
  Simon J.\ Puglisi\inst{3,4} \and
  Jouni Sir\'en\inst{5}}
\authorrunning{Cox et al.}  
\institute{Illumina Cambridge Ltd, UK \and
  University of Pisa, Italy \and
  Helsinki Institute for Information Technology, Finland \and
  University of Helsinki, Finland \and
  Wellcome Trust Sanger Institute, UK}
\maketitle

\begin{abstract}
Relative Lempel-Ziv (\rlz{}) is a popular algorithm for compressing databases of genomes from individuals of the same species when fast random access is desired.  With Kuruppu et al.'s (SPIRE 2010) original implementation, a reference genome is selected and then the other genomes are greedily parsed into phrases exactly matching substrings of the reference.  Deorowicz and Grabowski ({\it Bioinformatics}, 2011) pointed out that letting each phrase end with a mismatch character usually gives better compression because many of the differences between individuals' genomes are single-nucleotide substitutions.  Ferrada et al.\ (SPIRE 2014) then pointed out that also using relative pointers and run-length compressing them usually gives even better compression.  In this paper we generalize Ferrada et al.'s idea to handle well also short insertions, deletions and multi-character substitutions.  We show experimentally that our generalization achieves better compression than Ferrada et al.'s implementation with comparable random-access times.
\end{abstract}

\section{Introduction}
\label{sec:introduction}

Next-generation sequencing technologies can quickly and cheaply yield far more genetic data than can fit into an everyday computer's memory, so it is important to find ways to compress it while still supporting fast random access.  Often the data is highly repetitive and can thus be compressed very well with LZ77~\cite{ZL77}, but then random access is slow.  For many applications, however, we need store only a database of genomes from individuals of the same species, which are not only highly repetitive collectively but also but also all very similar to each other.  Kuruppu, Puglisi and Zobel~\cite{KPZ10} proposed choosing one of the genomes as a reference and then greedily parsing each of the others into phrases exactly matching substrings of that reference.  They called their algorithm Relative Lempel-Ziv (\rlz{}) because it can be viewed as a version of LZ77 that looks for phrase sources only in the reference, which greatly speeds up random access later.  (Ziv and Merhav~\cite{ZM93} introduced a similar algorithm for estimating the relative entropy of the sources of two sequences.)  RLZ is now is popular for compressing not only such genomic databases but also other kinds of repetitive datasets; see, e.g.,~\cite{HPZ11a,HPZ11b}.   Deorowicz and Grabowski~\cite{DG11} pointed out that letting each phrase end with a mismatch character usually gives better compression on genomic databases because many of the differences between individuals' genomes are single-nucleotide substitutions, and gave a new implementation with this optimization.  Ferrada, Gagie, Gog and Puglisi~\cite{FGGP14} then pointed out that often the current phrase's source ends two characters before the next phrase's source starts, so the distances between the phrases' starting positions and their sources' starting positions are the same.  They showed that using relative pointers and run-length compressing them usually gives even better compression on genomic databases.

In this paper we generalize Ferrada et al.'s idea to handle well also short insertions, deletions and substitutions.  In the Section~\ref{sec:preliminaries} we review in detail \rlz{} and Deorowicz and Grabowski's and Ferrada et al.'s optimizations.  We also discuss how \rlz{} can be used to build relative data structures and why the optimizations that work to better compress genomic databases fail for this application.  In Section~\ref{sec:adaptive} we explain the design and implementation of \rlz{} with adaptive pointers (\rlzap{}): in short, after parsing each phrase, we look ahead several characters to see if we can start a new phrase with a similar relative pointer; if so, we store the intervening characters as mismatch characters and store the new relative pointer encoded as its difference from the previous one.  We present our experimental results in Section~\ref{sec:experiments}, showing that \rlzap{} achieves better compression than Ferrada et al.'s implementation with comparable random-access times.  Finally, in Section~\ref{sec:future} we discuss directions for future work.  Our implementation and datasets are available for download from \mbox{\url{http://github.com/farruggia/rlzap}\ .}

\section{Preliminaries}
\label{sec:preliminaries}

In this section we discuss the previous work that is the basis and motivation for this paper.  We first review in greater detail Kuruppu et al.'s implementation of \rlz{} and Deorowicz and Grabowski's and Ferrada et al.'s optimizations.  We then quickly summarize the new field of {\em relative data structures} --- which concerns when and how we can use compress a new instance of a data structure, using an instance we already have for a similar dataset --- and explain how it uses \rlz{} and why it needs a generalization of Deorowicz and Grabowski's and Ferrada et al.'s optimizations.

\subsection{RLZ}
\label{subsec:rlz}

To compute the \rlz{} parse of a string \(S [1..n]\) with respect to a reference string $R$ using Kuruppu et al.'s implementation, we greedily parse $S$ from left to right into phrases
\begin{eqnarray*}
&& S [p_1 = 1..p_1 + \ell_1 - 1]\\
&& S [p_2 = p_1 + \ell_1..p_2 + \ell_2 - 1]\\
&& \hspace{12ex} \vdots\\
&& S [p_t = p_{t - 1} + \ell_{t - 1}..p_t + \ell_t - 1 = n]
\end{eqnarray*}
such that each \(S [p_i..p_i + \ell_i - 1]\) exactly matches some substring \(R [q_i..q_i + \ell_i - 1]\) of $R$ --- called the $i$th phrase's {\em source} --- for \(1 \leq i \leq t\), but \(S [p_i..p_i + \ell_i]\) does not exactly match any substring in $R$ for \(1 \leq i \leq t - 1\).  For simplicity, we assume $R$ contains every distinct character in $S$, so the parse is well-defined.

Suppose we have constant-time random access to $R$.  To support constant-time random access to $S$, we store an array \(Q [1..t]\) containing the starting positions of the phrases' sources, and a compressed bitvector \(B [1..n]\) with constant query time (see, e.g.,~\cite{KKP14} for a a discussion) and 1s marking the first character of each phrase.  Given a position $j$ between 1 and $n$, we can compute in constant time
\[S [i] = R \left[ \rule{0ex}{2ex} Q [B.\rank (j)] + j - B.\select (B.\rank (j)) \right]\,.\]
If there are few phrases then $Q$ is small and $B$ is sparse, so we use little space.

For example, if
\begin{eqnarray*}
R & = & \mathsf{ACATCATTCGAGGACAGGTATAGCTACAGTTAGAA}\\
S & = & \mathsf{ACATGATTCGACGACAGGTACTAGCTACAGTAGAA}
\end{eqnarray*}
then we parse $S$ into
\[\mathsf{ACAT}, \mathsf{GA}, \mathsf{TTCGA}, \mathsf{CGA}, \mathsf{CAGGTA}, \mathsf{CTA}, \mathsf{GCTACAGT}, \mathsf{AGAA}\,,\]
and store
\begin{eqnarray*}
Q & = & 1, 10, 7, 9, 15, 24, 23, 32\\
B & = & 10001010000100100000100100000001000\,.
\end{eqnarray*}
To compute \(S [25]\), we compute \(B.\rank (25) = 7\) and \(B.\select (7) = 24\), which tell us that \(S [25]\) is \(25 - 24 = 1\) character after the initial character in the 7th phrase.  Since \(Q [7] = 23\), we look up \(S [25] = R [24] = \mathsf{C}\).

\subsection{GDC}
\label{subsec:gdc}

Deorowicz and Grabowski~\cite{DG11} pointed out that with Kuruppu et al.'s implementation of \rlz{}, single-character substitutions usually cause two phrase breaks: e.g., in our example \(S [1..11] = \mathsf{ACATGATTCGA}\) is split into three phrases, even though the only difference between it and \(R [1..11]\) is that \(S [5] = \mathsf{G}\) and \(R [5] = \mathsf{C}\).  They proposed another implementation, called the Genome Differential Compressor (GDC), that lets each phrase end with a mismatch character --- as the original version of LZ77 does --- so single-character substitutions usually cause only one phrase break.  Since many of the differences between individuals' DNA are single-nucleotide substitutions, GDC usually compresses genomic databases better than Kuruppu et al.'s implementation.

Specifically, with GDC we parse $S$ from left to right into phrases \(S [p_1..p_1 + \ell_1], S [p_2 = p_1 + \ell_1 + 1..p_2 + \ell_2], \ldots, S [p_t = p_{t - 1} + \ell_{t - 1} + 1..p_t + \ell_t = n]\) such that each \(S [p_i..p_i + \ell_i - 1]\) exactly matches some substring \(R [q_i..q_i + \ell_i - 1]\) of $R$ --- again called the $i$th phrase's source --- for \(1 \leq i \leq t\), but \(S [p_i..p_i + \ell_i]\) does not exactly match any substring in $R$, for \(1 \leq i \leq t - 1\).

Suppose again that we have constant-time random access to $R$.  To support constant-time random access to $S$, we store an array \(Q [1..t]\) containing the starting positions of the phrases' sources, an array \(M [1..t]\) containing the last character of each phrase, and a compressed bitvector \(B [1..n]\) with constant query time and 1s marking the last character of each phrase.  Given a position $j$ between 1 and $n$, we can compute in constant time
\[S [j] = \left\{ \begin{array}{l@{\hspace{2ex}}l}
M [B.\rank (j)] & \mbox{if \(B [j] = 1\),}\\[1ex]
R \left[ \rule{0ex}{2ex} Q [B.\rank (j) + 1] + j - B.\select(B.\rank (j)) - 1 \right] & \mbox{otherwise,}
\end{array} \right.\]
assuming \(B.\select (0) = 0\).

In our example, we parse $S$ into
\[\mathsf{ACATG}, \mathsf{ATTCGAC}, \mathsf{GACAGGTAC}, \mathsf{TAGCTACAGT}, \mathsf{AGAA}\,,\]
and store
\begin{eqnarray*}
Q & = & 1, 6, 13, 21, 32\\
M & = & \mathsf{GCCTA}\\
B & = & 00001000000100000000100000000010001\,.
\end{eqnarray*}
To compute \(S [25]\), we compute \(B [25] = 0\), \(B.\rank (25) = 3\) and \(B.\select (3) = 21\), which tell us that \(S [25]\) is \(25 - 21 - 1 = 3\) characters after the initial character in the 4th phrase.  Since \(Q [4] = 21\), we look up \(S [25] = R [24] = \mathsf{C}\).

\subsection{Relative pointers}
\label{subsec:pointers}

Ferrada, Gagie, Gog and Puglisi~\cite{FGGP14} pointed out that after a single-character substitution, the source of the next phrase in GDC's parse often starts two characters after the end of the source of the current phrase: e.g., in our example the source for \(S [1..5] = \mathsf{ACATG}\) is \(R [1..4] = \mathsf{ACAT}\) and the source for \(S [6..12] = \mathsf{ATTCGAC}\) is \(R [6..11] = \mathsf{ATTCGA}\).  This means the distances between the phrases' starting positions and their sources' starting positions are the same.  They proposed an implementation of \rlz{} that parses $S$ like GDC does but keeps a relative pointer, instead of the explicit pointer, and stores the list of those relative pointers run-length compressed.  Since the relative pointers usually do not change after single-nucleotide substitutions, \rlz{} with relative pointers usually gives even better compression than GDC on genomic databases.  (We note that Deorowicz, Danek and Niemiec~\cite{DDN15} recently proposed a new version of GDC, called GDC2, that has improved compression but does not support fast random access.)

Suppose again that we have constant-time random access to $R$.  To support constant-time random access to $S$, we store the array $M$ of mismatch characters and the bitvector $B$ as with GDC.  Instead of storing $Q$, we build an array \(D [1..t]\) containing, for each phrase, the difference \(q_i - p_i\) between its source's starting position and its own starting position.  We store $D$ run-length compressed: i.e., we partition it into maximal consecutive subsequences of equal values, store an array $V$ containing one copy of the value in each subsequence, and a bitvector \(L [1..t]\) with constant query time and 1s marking the first value of each subsequence.  Given $k$ between 1 and $t$, we can compute in constant time
\[D [k] = V [L.\rank (k)]\,.\]
Given a position $j$ between 1 and $n$, we can compute in constant time
\[S [j] = \left\{ \begin{array}{l@{\hspace{2ex}}l}
M [B.\rank (j)] & \mbox{if \(B [j] = 1\),}\\[1ex]
R \left[ \rule{0ex}{2ex} D [B.\rank (j) + 1] + j \right]& \mbox{otherwise.}
\end{array} \right.\]

In our example, we again parse $S$ into
\[\mathsf{ACATG}, \mathsf{ATTCGAC}, \mathsf{GACAGGTAC}, \mathsf{TAGCTACAGT}, \mathsf{AGAA}\,,\]
and store
\begin{eqnarray*}
M & = & \mathsf{GCCTA}\\
B & = & 00001000000100000000100000000010001\,,
\end{eqnarray*}
but now we store \(D = 0, 0, 0, -1, 0\) as \(V = 0, -1, 0\) and \(L = 10011\) instead of storing $Q$.  To compute \(S [25]\), we again compute \(B [25] = 0\) and \(B.\rank (25) = 3\), which tell us that \(S [25]\) is in the 4th phrase.  We add 25 to the 4th relative pointer \(D [4] = V [L.\rank (4)] = -1\) and obtain 24, so \(S [25] = R [24]\).

A single-character insertion or deletion usually causes only a single phrase break in the parse but a new run in $D$, with the values in the run being one less or one more than the values in the previous run.  In our example, the insertion of \(S [21] = \mathsf{C}\) causes the value to decrement to -1, and the deletion of \(R [26] = \mathsf{T}\) (or, equivalently, of \(R [27] = \mathsf{T}\)) causes the value to increment to 0 again.  In larger examples, where the values of the relative pointers are often a significant fraction of $n$, it seems wasteful to store a new value uncompressed when it differs only by 1 from the previous value.

For example, suppose $R$ and $S$ are thousands of characters long,
\begin{eqnarray*}
R [1783..1817] & = & \ldots\mathsf{ACATCATTCGAGGACAGGTATAGCTACAGTTAGAA}\ldots\\
S [2009..2043] & = & \ldots\mathsf{ACATGATTCGACGACAGGTACTAGCTACAGTAGAA}\ldots
\end{eqnarray*}
and GDC still parses \(S [2009..2043]\) into the same phrases as before, with their sources in \(R [1783..1817]\).  The relative pointers for those phrases are \(-136, -136,\) \(-136, -137, -136\), so we store \(-136, -137, -136\) for them in $V$, which takes at least a couple of dozen bits without further compression.

\subsection{Relative data structures}
\label{subsec:structures}

As mentioned in Section~\ref{sec:introduction}, the new field of relative data structures concerns when and how we can use compress a new instance of a data structure, using an instance we already have for a similar dataset.  Suppose we have a basic FM-index~\cite{FM05} for $R$ --- i.e., a rank data structure over the Burrows-Wheeler Transform (BWT)~\cite{BW94} of $R$, without a suffix-array sample --- and we want to use it to build a very compact basic FM-index for $S$.  Since $R$ and $S$ are very similar, it is not surprising that their BWTs are also fairly similar:
\begin{eqnarray*}
\BWT (R) & = & \mathsf{AAGGT\$TTGCCTCCAAATTGAGCAAAGACTAGATGA}\\
\BWT (S) & = & \mathsf{AAGGT\$GTTTCCCGAAAATGAACCTAAGACGGCTAA}\,.
\end{eqnarray*}
Belazzougui, Gog, Gagie, Manzini and Sir\'en~\cite{BGGMS14} (see also~\cite{BBGMS15}) showed how we can implement such a relative FM-index for $S$ by choosing a common subsequence of the two BWTs and then storing bitvectors marking the characters not in that common subsequence, and rank data structures over those characters.  They also showed how to build a relative suffix-array sample to obtain a fully-functional relative FM-index for $S$, but reviewing that is beyond the scope of this paper.

An alternative to Belazzougui et al.'s basic approach is to compute the \rlz{} parse of \(\BWT (S)\) with respect to \(\BWT (R)\) and then store the rank for each character just before the beginning of each phrase.  We can then answer a rank query \(\BWT (S).\rank_X (j)\) by finding the beginning \(\BWT (S) [p]\) of the phrase containing \(\BWT (S) [j]\) and the beginning \(\BWT (R) [q]\) of that phrase's source, then computing
\[\BWT (S).\rank_X (p - 1) + \BWT (R).\rank_X (q + j - p) - \BWT (R).\rank_X (q - 1)\,.\]

Unfortunately, single-character substitutions between $R$ and $S$ usually cause insertions, deletions and multi-character substitutions between \(\BWT (R)\) and \(\BWT (S)\), so Deorowicz and Grabowski's and Ferrada et al.'s optimizations no longer help us, even when the underlying strings are individuals' genomes.  On the other hand, on average those insertions, deletions and multi-character substitutions are fairly few and short~\cite{LMS12}, so there is still hope that those optimized parsing algorithms can be generalized and applied to make this alternative practical.

Our immediate concern is with a recent implementation of relative suffix trees~\cite{GNPS15}, which uses relative FM-indexes and relatively-compressed longest-common-prefix (LCP) arrays.  Deorowicz and Grabowski's and Ferrada et al.'s optimizations also fail when we try to compress the LCP arrays, and when we use Kuruppu et al.'s implementation of \rlz{} the arrays take a substantial fraction of the total space.  In our example, however,
\begin{eqnarray*}
\LCP (R) & = & \mathsf{0,\!1,\!1,\!4,\!3,\!1,\!2,\!2,\!3,\!2,\!1,\!2,\!2,\!0,\!3,\!2,\!3,\!1,\!1,\!0,\!2,\!2,\!1,\!1,\!2,\!1,\!2,\!0,\!2,\!3,\!2,\!1,\!2,\!1,\!2}\\
\LCP (S) & = & \mathsf{0,\!1,\!1,\!4,\!3,\!2,\!2,\!1,\!2,\!2,\!2,\!1,\!2,\!0,\!3,\!2,\!1,\!4,\!1,\!3,\!0,\!2,\!3,\!2,\!1,\!1,\!1,\!3,\!0,\!3,\!2,\!3,\!1,\!1,\!1}
\end{eqnarray*}
are quite similar: e.g., they have a common subsequence of length 26, almost three quarters of their individual lengths.  LCP values tend to grow at least logarithmically with the size of the strings, so good compression becomes more important.

\section{Adaptive Pointers}
\label{sec:adaptive}

We generalize Ferrada et al.'s optimization to handle short insertions, deletions and substitutions by introducing {\em adaptive pointers} and by allowing more than one mismatch character at the end of each phrase.  An adaptive pointer is represented as the difference from the previous non-adaptive pointer.  Henceforth we say a phrase is \emph{adaptive} if its pointer is adaptive, and \emph{explicit} otherwise.  In this section we first describe our parsing strategy and then describe how we can support fast random access.

\subsection{Parsing}
\label{subsec:parsing}

The parsing strategy is a generalization of the Greedy approach for adaptive phrases.
The parser first compute the \emph{matching statistics} between input $S$ and reference $R$: for each suffix $S[i;n]$ of $S$, a suffix $R[k;m]$ of $R$ with the longest \LCP{} with $S[i]$ is found. Let \MatchRelPtr{i} be the relative pointer $k - i$ and \MatchLen{i} be the length of the \LCP{} between the two suffixes $S[i;n]$ and $R[k;m]$.

Parsing scans $S$ from left to right, in one pass. Let us assume $S$ has already been parsed up to a position $i$, and let us assume the most recent explicit phrase starts at position $h$.
The parser first tries to find an adaptive phrase (\emph{adaptive step}); if it fails, looks for an explicit phrase (\emph{explicit step}). Specifically:
\begin{enumerate}
\item \emph{adaptive step}: the parser checks, for the current position $i$ if \begin{inparaenum}[(i)]\item the relative pointer $\MatchRelPtr{i}$ can be represented as an adaptive pointer, that is, if the differential $\MatchRelPtr{i} - \MatchRelPtr{j}$ can be represented as a signed binary integer of at most $\DeltaBits$ bits, and \item if it is convenient to start a new adaptive phrase instead of representing literals as they are, that is, whether $\MatchLen{i} \cdot \log \sigma > \DeltaBits{}$\end{inparaenum}. The parser outputs the adaptive phrase and advances $\MatchLen{i}$ positions if both conditions are satisfied; otherwise, it looks for the leftmost position $k$ in range $i + 1$ up to $i + \LookAhead$ where both conditions are satisfied. If it finds such position $k$, the parser outputs literals $S[i;k-1]$ and an adaptive phrase; otherwise, it goes to step~\ref{en:explicit-step}.
\item\label{en:explicit-step} \emph{explicit step}: in this step the parser goes back to position $i$ and scans forward until it has found a position $k \geq i$ where at least one of these two conditions is satisfied:\begin{inparaenum}[(i)]\item match length \MatchLen{i} is greater than a parameter $\MinExplicitLength{}$; \item the match is followed by an adaptive phrase\end{inparaenum}. It then outputs a literal range $S[i;k-1]$ and the explicit phrase found.
\end{enumerate}
The purpose of the two conditions on the explicit phrase is to avoid having spurious explicit phrases which are not associated to a meaningfully aligned substrings.

It is important to notice that our data structure logically represents an adaptive/explicit phrase followed by a literal run as a single phrase: for example, an adaptive phrase of length $5$ followed by a literal sequence $\mathsf{GAT}$ is represented as an adaptive phrase of length $8$ with the last $3$ symbols represented as literals.

\subsection{Representation}
\label{subsec:representation}

In order to support fast random access to $S$, we deploy several data structures, which can be grouped into two sets with different purposes:
\begin{enumerate}
\item \textbf{Storing the parsing}: a set of data structures mapping any position $i$ to some useful information about the phrase $P_i$ containing $S[i]$, that is:\begin{inparaenum}[(i)]\item the position $\Start{i}$ of the first symbol in $P_i$; \item $P_i$'s length $\Length{i}$; \item its relative pointer $\Rel{i}$; \item the number of phrases $\PrevPhr{i}$ preceding $P_i$ in the parsing, and \item the number of explicit phrases $\AbsPhr{i} \geq \PrevPhr{i}$ preceding $P_i$.\end{inparaenum}
\item \textbf{Storing the literals}: a set of data structures which, given a position $i$ and the information about phrase $P_i$, tells whether $S[i]$ is a literal in the parsing and, if this is the case, returns $S[i]$.
\end{enumerate}
Here we provide a detailed illustration of these data structures.

\vspace{-1.5ex}

\paragraph{Storing the parsing.}
The parsing is represented by storing two bitvectors. The first bitvector $\PhraseBV{}$ has $|S|$ entries, marking with a $1$ characters in $S$ at the beginning of a new phrase in the parsing. The second bitvector $\ExplicitBV{}$ has $m$ entries, one for every phrases in the parsing, and marks every explicit phrase in the parsing with a $1$, otherwise $0$.
A rank/select datastructure is built on top of $\PhraseBV{}$, and a rank datastructure on top of $\ExplicitBV{}$. In this way, given $i$ we can efficiently compute the phrase index $\PrevPhr{i}$ as $\PhraseBV{}.rank{}(i)$, the explicit phrase index $\AbsPhr{i}$ as $\ExplicitBV{}.rank(p_i)$ and the phrase beginning $\Start{i}$ as $\PhraseBV{}.\select{}(p_i)$.

Experimentally, bitvector \PhraseBV{} is sparse, while \ExplicitBV{} is usually dense. Bitvector \PhraseBV{} can be represented with any efficient implementation for sparse bitvectors; our implementation, detailed in Section~\ref{sec:experiments}, employs the Elias-Fano based \textsf{SDarrays} datastructure of Okanohara and Sadakane \cite{OS07}, which requires $m \log \frac{|S|}{m} + O(m)$ bits and supports rank in $O(\log \frac{|S|}{m})$ time and select in constant time. Bitvector \ExplicitBV{} is represented plainly, taking $m$ bits, with any $o(m)$-space $O(1)$-time rank implementation on top of it (\cite{OS07,RRR07}). In particular, it is interesting to notice that only one $\rank{}$ query is needed for extracting an unbounded number of consecutive symbols from \ExplicitBV{}, since each starting position of consecutive phrases can be accessed with a single $\select{}$ query, which has very efficient implementations on sparse bitvectors.

Both explicit and relative pointers are stored using minimal binary codes in tables $A$ and $R$, respectively. These integers are not compressed using statistical encoding because this would prevent efficient random access to the sequence.
Each explicit and relative pointer takes thus $\lceil \log n \rceil$ and $\lceil \log{}(\LookAhead{}) \rceil + 1$ bits of space, respectively.
To compute $\Rel{i}$, we first check if the phrase is explicit by checking if $\mathsf{S}[\AbsPhr{i}]$ is set to one; if it is, then $\Rel{i} = A[\AbsPhr{i}]$, otherwise it is $\Rel{i} = A[\AbsPhr{i}] + R[\PrevPhr{i} - \AbsPhr{i}]$.

\vspace{-1.5ex}

\paragraph{Storing literals.}
Literals are extracted as follows. Let us assume we are interested in accessing $S[i]$, which is contained in phrase $P_j$. First, it is determined whether $S[i]$ is a literal or not. Since literals in a phrase are grouped at the end of the phrase itself, it is sufficient to store, for every phrase $P_k$ in the parsing, the number of literals $\Literals{k}$ at its end. Thus, knowing the starting position $\Start{j}$ and length $\Length{j}$ of phrase $P_j$, symbol $S[i]$ is a literal if and only if $i > \Start{j} + \Length{j} - \Literals{j}$.

All literals are stored in a table $L$, where $L[k]$ is the $k$-th literal found by scanning the parsing from left to right.  How we represent $L$ depends on the kind of data we are dealing with.  In our experiments, described in Section~\ref{sec:experiments}, we consider differentially-encoded LCP arrays and DNA.  For \textsf{DLCP} values, $L$ simply stores all values using minimal binary codes. For \textsf{DNA} values, a more refined implementation (which we describe in a later paragraph) is needed to use less than $3$ bits on average for each symbol.
So, in order to display the literal $S[i]$, we need a way to compute its index in $L$, which is equal to $\Start{j} - \Length{j} - \Literals{k}$ plus the prefix sum $\sum_{k = 1}^{j-1} \Literals{k}$. In the following paragraph we detail two solutions for efficiently storing \Literals{k} values and computing prefix sums.

\vspace{-1.5ex}

\paragraph{Storing literal counts.}
Here we detail a simple and fast data structure for storing \Literals{-} values and for computing prefix sums on them.
The basic idea is to store \Literals{-} values explicitly, and accelerate prefix sums by storing the prefix sum of some regularly sampled positions. To provide fast random access, the maximum number of literals in a phrase is limited to $2^\MaxLit{} - 1$, where $\MaxLit{}$ is a parameter chosen at construction time. Every value $\Literals{-}$ is thus collected in a table $L$, stored using $\MaxLit{}$ bits each.
Since each phrase cannot have more than $2^\MaxLit{} - 1$ literals, we split each run of more than $2^\MaxLit{} - 1$ literals into the minimal number of phrases which do meet the limit.  In order to speed-up the prefix sum computation on $L$, we sample one every \SampleInterval{} positions and store prefix sums of sampled positions into a table \Prefix{}. To accelerate further prefix sum computation, we employ a $256$-entries table \StaticTable{} which maps any sequence of $8 / \MaxLit{}$ elements into their sum. Here, we constrain $\MaxLit{}$ as a power of two not greater than $8$ (that is, either $1$, $2$, $4$ or $8$) and \SampleInterval{} as a multiple of $8 / \MaxLit{}$. In this way we can compute the prefix sum by just one look-up into \Prefix{} and at most $\frac{\SampleInterval{}}{8 / \MaxLit{}}$ queries into \StaticTable{}. Using \StaticTable{} is faster than summing elements in $L$ because it replaces costly bitshift operations with efficient byte-accesses to $L$. This is because $8 / \MaxLit{}$ elements of $L$ fit into one byte; moreover, those bytes are aligned to byte-boundaries because $\SampleInterval{}$ is a multiple of $8 / \MaxLit{}$, which in turn implies that the sampling interval spans entire bytes of $L$.

\vspace{-1.5ex}

\paragraph{Storing DNA literals.}
Every literal is collected into a table $J$, where each element is represented using a fixed number of bits.
For the DNA sequences we consider in our experiments, this would imply using $3$ bits, since the alphabet is $\{A, C, G, T, N\}$. However, since symbols $N$ occur less often than the others, it is more convenient to handle those as exceptions, so other literals can be stored in just $2$ bits. In particular, every $N$ in table $J$ is stored as one of the other four symbols in the alphabet (say, $A$) and a bit-vector \NExc{} marks every position in $J$ which corresponds to an $N$. Experimentally, bitvector \NExc{} is sparse and the $1$ are usually clustered together into a few regions. In order to reduce the space needed to store \NExc{}, we designed a simple bit-vector implementation to exploit this fact.
In our design, \NExc{} is divided into equal-sized chunks of length $C$. A bitvector $\mathsf{Chunk}$ marks those chunks which contain at least one bit set to $1$. Marked chunks of \NExc{} are collected into a vector $V$. Because of the clustering property we just mentioned, most of the chunks are not marked, but marked chunks are locally dense. Because of this, bitvector $\mathsf{Chunk}$ is implemented using a sparse representation, while each chunk employs a dense representation. Good experimental values for $C$ are around $16-32$ bits, so each chunk is represented with a fixed-width integer. In order to check whether a position $i$ is marked in \NExc{}, we first check if chunk $c = \lfloor i / C \rfloor$ is marked in $\mathsf{Chunk}$. If it is marked, we compute $\mathsf{Chunk}.rank(c)$ to get the index of the marked chunk in $V$.

\section{Experiments}
\label{sec:experiments}

We implemented \rlzap{} in C++11 with bitvectors from Gog et al.'s \textsf{sdsl} library \mbox{(\url{https://github.com/simongog/sdsl-lite})}, and compiled it with \verb=gcc= version \verb=4.8.4= with flags \verb|-O3|, \verb|-march=native|, \verb|-ffast-math|, \verb|-funroll-loops| and \verb|-DNDEBUG|.  We performed our experiments on a computer with a $6$-core Intel Xeon X5670 clocked at 2.93GHz, $40$GiB of DDR3 ram clocked at 1333MHz and running Ubuntu 14.04.  As noted in Section~\ref{sec:introduction}, our code is available at \mbox{\url{http://github.com/farruggia/rlzap}\ .}

We performed our experiments on the following four datasets:
\begin{itemize}
\item \textsf{Cere}: the genomes of $39$ strains of the \emph{Saccharomyces cerevisiae} yeast;
\item \textsf{E.\@ Coli}: the genomes of $33$ strains of the \emph{Escherichia coli} bacteria;
\item \textsf{Para}: the genomes of $36$ strains of the \emph{Saccharomyces paradoxus} yeast;
\item \textsf{DLCP}: differentially-encoded LCP arrays for three human genomes, with 32-bit entries.
\end{itemize}
These files are available from \url{http://acube.di.unipi.it/rlzap-dataset}.

For each dataset we chose the file (i.e., the single genome or DLCP array) with the lexicographically largest name to be the \emph{reference}, and made the concatenation of the other files the \emph{target}.  We then compressed the target against the reference with Ferrada et al.'s optimization of \rlz{} --- which reflects the current state of the art, as explained in Section~\ref{sec:introduction} --- and with \rlzap{}.  For the DNA files (i.e., \textsf{Cere}, \textsf{E.\@ Coli} and \textsf{Para}) we used $\LookAhead{} = 32$, $\MinLength{} = 32$ and $\DeltaBits{} = 2$, while for \textsf{DLCP} we used $\LookAhead{} = 8$, $\MinLength{} = 4$ and $\DeltaBits{} = 4$.  We chose these parameters during a calibration step performed on a different dataset, which we will describe in the full version of this paper.

Table~\ref{tab:compression} shows the compression achieved by \rlz{} and \rlzap{}.  (We note that, since the DNA datasets are each over an alphabet of \(\{\mathsf{A}, \mathsf{C}, \mathsf{G}, \mathsf{T}, \mathsf{N}\}\) and {\sf N}s are rare, the targets for those datasets can be compressed to about a quarter of their size even with only, e.g., Huffman coding.)  Notice \rlzap{} consistently achieves better compression than \rlz{}, with its space usage ranging from about 17\% less for \textsf{Cere} to about 32\% less for \textsf{DLCP}.

\begin{table}[t]
\caption{Compression achieved by \rlz{} and \rlzap{}.  For each dataset we report in MiB ($2^{20}$ bytes) the size of the reference and the size of the target uncompressed and compressed with each method.}
\label{tab:compression}
\begin{center}
\begin{tabular}{l @{\hspace{3ex}} r @{\hspace{3ex}} r @{\hspace{3ex}} r @{\hspace{8ex}} r @{\hspace{3ex}} r @{\hspace{6ex}}}
\toprule
Dataset  				    & \multicolumn{1}{l}{Reference}	& \multicolumn{1}{l}{Target}		& \multicolumn{3}{l}{Compressed Target Size (MiB)} \\
						        & size (MiB)             & size (MiB)          & \rlz{}     & \rlzap{}  \\
\midrule
\textsf{Cere}       & 12.0                   & 451                 & 9.16       & 7.61      \\
\textsf{E. \@Coli}  & 4.8                    & 152                 & 30.47      & 21.51     \\
\textsf{Para}       & 11.3                   & 398                 & 15.57      & 10.49     \\
\textsf{DLCP}       & 11,582                 & 23,392              & 1,745.33   & 1,173.81  \\
\bottomrule
\end{tabular}
\end{center}
\end{table}

\begin{table}[t]
\caption{Extraction times per character from \rlz{}- and \rlzap{}-compressed targets.  For each file in each target, we compute the mean extraction time for $2^{24} / \ell$ pseudo-randomly chosen substrings; take the mean of these means.}
\label{tab:decompression_time}
\begin{center}
\begin{tabular}{C{5em} C{6em} R{3em} R{3em} R{3em} R{3em} R{3em} R{3em}}
\toprule
\multirow{2}*{Dataset}            	&  \multirow{2}*{Algorithm}	& \multicolumn{6}{c}{Mean extraction time per character (ns)}\\
                                   &           & 1    &  4   & 16    &   64  &  256   & 1024  \\
\midrule
\multirow{2}*{\textsf{Cere}}       & \rlz{}    & 234  & 59   & 16.4  & 4.4   & 1.47   & 0.55  \\
                                   & \rlzap{}  & 274  & 70   & 19.5  & 5.7   & 2.34   & 1.26  \\
\midrule
\multirow{2}*{\textsf{E. \@Coli}}  & \rlz{}    & 225  & 62   & 20.1  & 7.7   & 4.34   & 3.34  \\
                                   & \rlzap{}  & 322  & 91   & 31.3  & 15.3  & 10.78  & 9.47  \\
\midrule
\multirow{2}*{\textsf{Para}}       & \rlz{}    & 235  & 59   & 17.2  & 5.2   & 2.23   & 1.03  \\
                                   & \rlzap{}  & 284  & 74   & 21.2  & 6.9   & 3.09   & 2.26  \\
\midrule
\multirow{2}*{\textsf{DLCP}}       & \rlz{}    & 756  & 238  & 61.5  & 20.5  & 9.00   & 6.00  \\
                                   & \rlzap{}  & 826  & 212  & 57.5  & 19.0  & 8.00   & 4.50  \\
\bottomrule
\end{tabular}
\end{center}
\end{table}

Table~\ref{tab:decompression_time} shows extraction times for \rlz{}- and \rlzap{}-compressed targets.  \rlzap{} is noticeably slower than \rlz{} for DNA, while it is slightly faster for the DLCP dataset when at least four characters are extracted.  We believe \rlzap{} outperforms \rlz{} on the DLCP because its parsing is generally more cache-friendly: our measurements indicate that on this dataset \rlzap{} causes about 36\% fewer L2 and L3 cache misses than \rlz{}.  Even for DNA, \rlzap{} is still fast in absolute terms, taking just tens of nanoseconds per character when extracting at least four characters.

On DNA files, \rlzap{} achieves better compression at the cost of slightly longer extraction times.  On differentially-encoded LCP arrays, \rlzap{} outperforms \rlz{} in all regards, except for a slight slowdown when extraction substrings of length less than 4.  That is, \rlzap{} is competitive with the state of the art even for compressing DNA and, as we hoped, advances it for relative data structures.  Our next step will be to integrate it into the implementation of relative suffix trees mentioned in Subsection~\ref{subsec:structures}.

\section{Future Work}
\label{sec:future}

In the near future we plan to perform more experiments to tune \rlzap{} and discover its limitations.  For example, we will test it on the balanced-parentheses representations of suffix trees' shapes, which are an alternative to LCP arrays, and on the BWTs in relative FM-indexes.  We also plan to investigate how to minimize the bit-complexity of our parsing --- i.e., how to choose the phrases and sources so as to minimize the number of bits in our representation --- building on the results by Farruggia, Ferragina and Venturini~\cite{FFV14a,FFV14b} about minimizing the bit-complexity of LZ77.

\rlzap{} can be viewed as a bounded-lookahead greedy heuristic for computing a glocal alignment~\cite{BMPDCDB03} or $S$ against $R$.  Such an alignment allows for genetic recombination events, in which potentially large sections of DNA are rearranged.  We note that standard heuristics for speeding up edit-distance computation and global alignment do not work here, because even a low-cost path through the dynamic programming matrix  can occasionally jump arbitrarily far from the diagonal.  \rlzap{} runs in linear time, which is attractive, but it may produce a suboptimal alignment --- i.e., it is not an admissible heuristic.  In the longer term, we are interested in finding practical admissible heuristics.

For example, if a long enough substring of $S$ aligns well enough against a particular substring of $R$ and badly enough against any other substring or small collection of substrings of $R$ (which we can check with LCP queries), then any optimal alignment of $S$ against $R$ should align most of that subalignment.  This observation should help us find an optimal alignment when the \rlz{} parse of $S$ with respect to $R$ is small but, e.g., there are few or no long approximate repetitions within $R$, so the LZ77 parse of $R$ is fairly large.

Apart from the direct biological interest of computing optimal or nearly optimal glocal alignments, they can also help us design more data structures.  For example, consider the problem of representing the mapping between orthologous genes in several species' genomes; see, e.g.,~\cite{Kub14}.  Given two genomes' indices and the position of a base-pair in one of those genomes, we would like to return quickly the positions of all corresponding base-pairs in the other genome.  Only a few base-pairs correspond to two base-pairs in another genome and, ignoring those, this problem reduces to representing compressed permutations.  A feature of these permutations is that base-pairs tend to be mapped in blocks, possibly with some slight reordering within each block.  We can extract this block structure by computing a glocal alignment, either between the genomes or between the permutation and its inverse.

\enlargethispage{10ex}

\end{document}